\documentclass[twocolumn,prb,aps,showpacs]{revtex4}
\usepackage{graphics}% Include figure files
\usepackage{dcolumn}% Align table columns on decimal point
\usepackage{bm}% bold math
\usepackage{amsmath}
\usepackage{amssymb}
\usepackage{epsfig}
\usepackage{pictex}
\begin{document}
\title{ Metal-insulator transition in half-filled two-orbital
        Hubbard model on triangular lattice}
\author{Feng Lu$^{1,2}$,  Wei-Hua Wang$^{1,3}$, and Liang-Jian Zou$^{1
\footnote{Correspondence author, Electronic mail: zou@theory.issp.ac.cn}}$}
\affiliation{\it
 $^1$ Key Laboratory of Materials Physics, Institute of Solid State Physics,
      Chinese Academy of Sciences, P. O. Box 1129, Hefei 230031, China  \\
\it $^2$  Graduate School of the Chinese Academy of Sciences,
          Beijing 100049, China \\
\it $^3$ Department of Electronics, College of Information Technical
         Science, Nankai University, Tianjin 300071, China}
\date{2008-2-14}

\begin{abstract}

We have investigated the half-filled two-orbital Hubbard model on a
triangular lattice by means of the dynamical mean-field theory
(DMFT). The local squared moments of charge, spin and orbital, and
the optical conductivity clearly show that the metal-insulator
transition (MIT) occurs at U$_{c}$, U$_{c}$=18.2, 16.8, 6.12 and
5.85 for the Hund's coupling J=0, 0.01U, U/4 and U/3, respectively.
The distinct continuities of the double occupation of electrons, the
local squared moments and the local susceptibility of charge, spin
and orbital suggest that for J $>$ 0, the MIT is first-order;
however at J=0, the MIT is second-order.
We attribute the first-order nature of the MIT to the symmetry
lowering of the systems with finite Hund's coupling.

\end{abstract}

\pacs{71.30.+h, 64.60.-i, 78.20.-e, 71.27.+a}

\maketitle

\section{Introduction}
%
%( MIT, DMFT, order of MIT in single-orbital Hubbard model)
%
 Metal-Insulator transitions (MIT) and related properties in
correlated electron systems have been the central topics in
condensed matter physics for several decades \cite{MIT}. The MIT can
be easily realized by the variation of the external fields, doping
concentration, pressure and temperature in many typical
transition-metal oxides.
The simplest and effective model to describe the low-energy physics
of these strongly correlated transition-metal oxides is the
single-orbital Hubbard model, including the competition between the
kinetic energy and the local Coulomb interaction. Such a competition
may result in many complicated and novel phenomena, such as the high
temperature superconductivity in low-dimensional cuprates.
Theoretically, great progress has been achieved in understanding the
essence of the MIT in the single-orbital Hubbard model, mainly due
to the development of the dynamical mean-field theory
(DMFT)\cite{Georges} in the past decade.
The DMFT allows us to accurately treat the Hubbard subbands in time
axis and to obtain the quasiparticle peaks with the three-peak
structure, which makes this approach an advance over the density
functional theory and the Hatree-Fock approximation \cite{Kotliar}.
With the help of DMFT, we have gotten a deep insight to many
properties of the single-orbital Hubbard model, e.g. the MIT, the
optical conductivity and absorption, transport, and so
on\cite{Kotliar}.
Among these properties related to the MIT, the order of the MIT in
the Hubbard model is essential. Bulla {\it et al.}
\cite{Bulla,Bulla99} demonstrated that in the single-orbital Hubbard
model on Bethe lattice, the MIT is first-order for 0 $<$ T $<$
T$_{c}$; whilst, it is second-order for T $>$ T$_{c}$.
In the two-dimensional Hubbard model, Onoda and Imada \cite{Onoda}
also found that the MIT is first-order in finite T $<$ T$_{MIT}$ by
means of the correlator projection approach with the DMFT.
These suggest that in low T, the MIT in the single-orbital Hubbard
model is first-order.

%
%(2D triangular lattice, multi-orbital Hubbard model, complicated MIT)
%(NaNiO2, AgNiO2, NaMnO2...urgent study)
%
Since the realistic transition-metal oxides, such as manganites,
vanadates, titanates and nickelates, usually have multiple
degenerate orbitals \cite{Kugel}, the multi-orbital Hubbard model is
more appropriate to describe the low-energy process than the
singe-orbital Hubbard model. At the same time, the multi-orbital
Hubbard model may exhibit more complicated and richer phenomena than
the single-orbital Hubbard model. Besides the conventional
localization-delocalization transition of electrons, there may exist
many different orbital ordered phases. For example, in a two-orbital
system, one orbital may be completely empty and another is
fulfilled, forming the ferro/antiferro-orbital ordered phase; or,
one orbital is fulfilled and insulating, another is partially
occupied and metallic, forming the so-called orbital selective Mott
phase (OSMP) \cite{Anisimov, Koga04,Liebsch,Liebsch03,Liebsch04}. In
these situations, the orbital degree of freedom plays an important
role in the phase diagram and the groundstate properties.
%

%(DMFT on MIT of 2-orbital Hubbard model, present status & big problems)
%
More recently, a number of researches have been concentrated on the
nature of MIT and other properties of the two-orbital Hubbard models
\cite{Inaba05,Pruschke05,Inaba,Bunemann:Gutzwiller}.
However, even on the Bethe or the hyper-cubic lattices, the nature
of the MIT in the two-orbital Hubbard models has been controversial,
although intensive theoretical efforts have already been done
\cite{Bunemann:Gutzwiller,Inaba05,Pruschke05,Inaba}.
In the two-orbital systems, Inaba {\it et al.} \cite{Inaba05} and
B\"unemann {\it et al.} \cite{Bunemann:Gutzwiller} found that the
Mott transition is discontinuous for any finite $J>0$ and continuous
only for $J=0$ within a generalized Gutzwiller approximation.
However, utilizing the DMFT with the numerical renormalization
group, Pruschke and Bulla \cite{Pruschke05} claimed that the Mott
transition is second-order for J $>$ $U/4$. They found that the
variation of the local squared moment of spin near the transition is
too small to judge the order of MIT for large J.
By making use of the DMFT with self-energy functional approach,
Inaba and Koga \cite{Inaba} believed that the nature of the Mott
transition is first-order in all the parameter region for finite J,
though they found that the jump of quasiparticle weight is too weak
to identify the order of the phase transition when J is large
enough.
The controversy on the order of the MIT suggests that it is urgent
to find a more proper quantity to judge the occurrence and the order
of the MIT when J is very large.

Up to date, most of the studies have been focused on the Bethe or
the hyper-cubic lattices. It is not known what the essence of the
MIT is in the multi-orbital Hubbard model on the frustrated
lattices.
When the strong electron-electron interactions compete with the
geometrical frustration effects, a number of unconventional phases
and exotic properties emerge as the result of the competition, such
as the MIT and the antiferromagnetism in the organic compounds
$\kappa$-(BEDT-TTF)$_{2}$X with X as an anion
\cite{kanoda,mckenzie}, etc.
Recent development in material fabrication shows that more and more
transition-metal oxides exhibit strong electronic correlation on
two-dimensional triangular lattices and the multiple orbital
character, such as NaNiO$_{2}$\cite{Petit} and
AgNiO$_{2}$\cite{Coldea}, etc. These appeal for the study on the
multi-orbital Hubbard model on the triangular lattice.

In this paper we focus on the MIT physics of the two-orbital Hubbard
model on a triangular lattice by means of the exact-diagonalization
DMFT.
%
% Comparing with the Bethe lattices or hyper-cubic lattices in the
% previous studies, one expects that strongly correlated electrons on
% a two-dimensional triangular lattice suffer from strong frustration
% effect, and the antiferromagnetic long-range order is greatly
% suppressed in such a frustrated lattice, so the DMFT approach is
% suitable for exploring the nature of MIT in the two-orbital Hubbard
% models \cite{Aryanpour}.
%
We adopt the local squared moment of charge, together with the local
squared moments of spin and orbital, to measure the occurrence of
MIT in the two-orbital Hubbard model, and find that we can well
judge the occurrence of the MIT when the Hund's coupling J is very
large. We definitely show that the MIT at large J is first-order.
The variation of the optical conductivity of the two-orbital Hubbard
model is also consistent with the MIT with the increase of U.
The rest of this paper is organized as follows: in Sec.II, we
describe the model Hamiltonian of the two-orbital system and briefly
explain the framework of the exact-diagonalization DMFT approach; in
Sec.III, we present the evolutions of the densities of states (DOS),
the local squared moments of charge, spin and orbital, and the
optical conductivity with the on-site Coulomb interaction; the order
of the MIT in the two-orbital system is also discussed in Sec.III;
the last part is devoted to the summary.

\section{HAMILTONIAN AND METHOD}
\label{secmodel}

   We start from a half-filled two-orbital Hubbard model
%%%%%%%%%%%%%%%%%%%%%%%%%%%%%%%
\begin{eqnarray}
   H &=& \sum_{<i,j>, \alpha, \beta \sigma}%\sum_{\alpha}\sum_{\sigma}
         t_{\alpha\beta} c^\dag_{i\alpha\sigma} c_{j\beta\sigma},
          + \sum_i {H}_i^\prime
\\
   {H}_i^\prime&=& U \sum_{\alpha}
                n_{i \alpha \uparrow} n_{i \alpha \downarrow}
                  +U^\prime\sum_{\sigma \sigma^\prime}
                n_{i 1 \sigma} n_{i 2 \sigma^\prime}
                \nonumber\\
          &+&   J\sum_{\sigma \sigma^\prime}
                c^\dag_{1 \sigma}c^\dag_{2 \sigma^\prime}
                c_{ 1 \sigma^\prime}c_{ 2 \sigma}
                +J^\prime\sum_{\alpha \ne \beta}
                c^\dag_{\alpha \uparrow}c^\dag_{\alpha \downarrow}
                c_{\beta \downarrow}c_{\beta \uparrow}
          \label{eq:model},
\end{eqnarray}
%%%%%%%%%%%%%%%%%%%%%%%%%%%%%%%%%%%%%%%%%%
in a triangular lattice, where
$c^\dag_{i\alpha\sigma}(c_{i\alpha\sigma})$ is the creation
(annihilation) operator of the electron at site $i$ with orbital
$\alpha (=1,2)$ and spin $\sigma (=\uparrow, \downarrow)$, and
$n_{i\alpha\sigma}$ is the electron number operator.
$t_{\alpha\beta}$ denotes the hopping integral from the $\beta$
orbital to the $\alpha$ orbital, and only the nearest-neighbor
hopping is taken into account. For clarity and to compare our
results with the present literature, we assume that the
intra-orbital hopping integrals are the same, i.e.
$t_{\alpha\alpha}$ =$t_{\beta\beta}$ =t; and we neglect the
inter-orbital hopping, though sometimes the inter-orbital components
may play an important role\cite{Song05}.

The parameters $U$, $U^{\prime}$, $J$, and $J^{\prime}$ denote the
intra-orbital Coulomb, inter-orbital Coulomb, Hund's and the
pair-hopping couplings. In what follows, considering the realistic
wavefunctions of 3d-orbitals \cite{Castellani} and the spin
rotational symmetry, we adopt the relationships $J=J^{\prime}$ and
$U=U'+2J$.
Unlike the Bethe or hyper-cubic lattice, the particle-hole symmetry
is broken at half-filling on the triangular lattice. At U=0, the
tight-binding dispersion of each orbital channel is
\begin{equation}
   \epsilon_{\bf k\alpha\alpha}=-2t_{\alpha\alpha} [\cos(k_x) +2
             \cos\left({\sqrt{3} \over
           2} k_y\right) \cos\left({k_x \over 2}\right)], \label{disper}
\end{equation}
with the bandwidth $W=9|t|$.

%====================================================================
%==                             Method                             ==
%====================================================================
%\section{Method}\label{sec:method}
Within the framework of the DMFT, the Hamiltonian (1) and (2) are
mapped onto an effective Anderson impurity model by integrating over
all the spatial degrees of freedom, except for the central site $o$.
The corresponding Hamiltonian, $H_{eff}$, contains a central
"atomic" or "impurity" part, $H_{atom}$, and an effective medium
part $H_{med}$, which has to be determined self-consistently.  The
two-orbital Anderson impurity Hamiltonian reads
\begin{eqnarray}
  H_{\rm eff}
  & = &
 H_{\rm atom} + H_{\rm med}
 \: ,
\label{eq:mimp}
\end{eqnarray}
and
\begin{eqnarray}
  H_{\rm med}
  & = &
  \sum_{\alpha\sigma} \epsilon_{{\rm d}\alpha} \:
  d^\dagger_{\alpha\sigma} d_{\alpha\sigma}
\nonumber \\
  & + & \sum_{\alpha \sigma, k=2}^{n_{\rm s}}
  \epsilon_{k\alpha\sigma} \:
  a^\dagger_{k\alpha\sigma} a_{k\alpha\sigma}
\nonumber \\
  & + & \sum_{\alpha\sigma, k=2}^{n_{\rm s}}
  V_{k\alpha\sigma} \:
  (d^\dagger_{\alpha\sigma} a_{k\alpha\sigma} + \mbox{h.c.}) \: ,
\label{eq:mimp}
\end{eqnarray}
where $d^{\dag}_{\alpha\sigma}$ and $a^{\dag}_{\alpha\sigma}$ create
the "impurity" electron and a bath electron, respectively; the
impurity level $\epsilon_{{\rm d}\alpha}$ is usually chosen as the
zero point of energy, and the hybridization parameter $V_{k\alpha
\sigma}$ of the impurity model is calculated self-consistently in
DMFT. The atomic Hamiltonian $H_{atom}$ of the central site is the
same as $H^\prime$ in Eq.(2), and $n_{s}$ represents the number of
the conduction band of Anderson impurity model.
%
% The impurity model is solved self-consistently.
% The iterative procedure starts by solving the Anderson model for a given
% choice parameters. From the on-site Green's function,
% we may obtain the self-energy of the system
%
For a set of parameters ($U$ and $J$), we can obtain the interacting
Green function $G_{\alpha\sigma}(i \omega_n)$, the free Green
function $G_{0\alpha\sigma}(i\omega_n)$ and the self-energy of the
Anderson model as follows,
\begin{equation}
   \Sigma_{\sigma}^{\alpha \beta}(i\omega_n)=[G^{-1}_{0\sigma}(i
          \omega_n)-G^{-1}_{\sigma}(i \omega_n)]_{\alpha \beta},
\end{equation}
and the lattice Green's function is
\begin{eqnarray}
   G_\sigma^{\alpha \beta}(i \omega_n)&=&\sum_{\bf k} G_\sigma^{\alpha
       \beta}({\bf k},i\omega_n) \\
    &=&\sum_{\bf k} [{1 \over i \omega_n + \mu -\epsilon_{\bf
       k} -\Sigma_\sigma(i \omega_n)}]_{\alpha \beta},
\end{eqnarray}
and the impurity Green function
\(  G_{{\rm imp},\sigma}^{\alpha \beta}(i \omega) = \langle \langle
      d_{\alpha\sigma}; d_{\alpha\sigma}^\dagger \rangle \rangle_{i
      \omega} \)
is given by
\begin{equation}
  G_{{\rm imp},\sigma}^{\alpha \beta}(i \omega) =[
  \frac{1}{i \omega + \mu - \epsilon_{{\rm d}\alpha}
  - {\Delta}_{\sigma}(i \omega) - {\Sigma}_{{\rm imp},\sigma}(i
  \omega)}]_{\alpha \beta}
  \: .
\label{eq:mimpgreen}
\end{equation}
In Eq.(9), the spectral width function
$\Delta_{\alpha\sigma}(i\omega)=\sum_k V_{k\alpha\sigma}^2 /
(i\omega + \mu - \epsilon_{k\alpha\sigma})$, $\mu$ is the chemical
potential, and $\omega_n=(2n+1)\pi/\beta$ is the fermionic Matsubara
frequency. Throughout this paper we fix the temperature
$\beta=1/k_BT$=16. We perform the iterative procedure repeatedly
until a self-consistent solution of the lattice Green's function and
the self-energy are found.

Various analytical and numerical methods can be employed to solve the
effective impurity problem.
In the following, we first make use of the exact-diagonalization
(ED) method to treat the impurity model Eq.(4) and (5) at finite
temperatures of $\beta$ \cite{Georges}. Then we perform the
iteration on Eqs. (6-9) repeatedly until a self-consistent solution.
Some subroutines, such as the minimizing subroutine which searches
for the parameters of the Anderson impurity Hamiltonian and the RS
subroutine which is used to diagonalize the Anderson impurity
Hamiltonian, are from the Ref.[2].
In this paper we take $n_{s}=6$ for each spin channel. Liebsch found
that when $n_{s}$ $>$ 3, the converged results qualitatively agree
with each other \cite{Liebsch05}, and the accuracy of the ED ansatz
in the single-orbital Hubbard model is well controlled.
Demchenko $et$ $al.$ had shown that in the absence of particle-hole
symmetry, the pole formation and the MIT are independent of
each other on the Bethe lattice\cite{Demchenko}.
So, the quasiparticle weight Z is not suitable for characterizing
the occurrence of the MIT. In this paper, we utilize the local
squared moments of charge, spin and orbital and the corresponding local
susceptibility to characterize the nature of the Mott transition.

 Since the MIT is associated with
the localization-delocalization transition of the charge degree of
freedom, we measure the MIT with the local squared moments of
charge, $<C^{2}>$,
\begin{eqnarray}
   <C^{2}> &=&\langle (n-2)^{2} \rangle  \nonumber
\end{eqnarray}
together with the local squared moments of spin and orbital,
\begin{eqnarray}
  <S_{z}^{2}>&=&\langle (n_{\uparrow}-n_{\downarrow})^{2} \rangle
\nonumber
\end{eqnarray}
\begin{eqnarray}
  <T_{z}^{2}>&=&\langle (n_{1}-n_{2})^{2} \rangle.
\end{eqnarray}
All of these quantities are relevant to the spin-dependent double
occupancy $<n_{\uparrow}n_{\downarrow}>$ and the orbital-dependent
double occupancy $<n_{1}n_{2}>$:
\begin{eqnarray}
  <n_{\uparrow}n_{\downarrow}> &=& \frac{\partial F}{\partial U}
   =<n_{1\uparrow}n_{1\downarrow}+n_{2\uparrow}n_{2\downarrow}>,
\nonumber \\
  <n_{1}n_{2}> &=& \frac{\partial F}{\partial U^{\prime}}
  =<(n_{1\uparrow}+n{1\downarrow})(n_{2\uparrow}+n{2\downarrow})>,
\end{eqnarray}
through a few identities, such as,
\begin{eqnarray}
   <C^{2}>=\frac{\partial F}{\partial \mu}+2(\frac{\partial F}{\partial
          U}+\frac{\partial F}{\partial U^{\prime}}),
\end{eqnarray}
here $F$ is the free energy, and $\mu$ the chemical potential.
The local susceptibilities of the charge, spin and
orbital are defined by
\begin{eqnarray}
  \chi_c&=&\int^\beta_0 \langle {\cal T} [n(\tau)-2][n(0)-2]\rangle
           d\tau,
                   \nonumber\\
  \chi_s&=&\int^\beta_0 \langle {\cal T} [n_\uparrow(\tau)-
      n_\downarrow(\tau)][n_\uparrow(0)-n_\downarrow(0)]\rangle d\tau,
                    \nonumber\\
  \chi_o&=&\int^\beta_0 \langle {\cal T}
          [n_1(\tau)-n_2(\tau)][n_1(0)-n_2(0)]\rangle d\tau,
\label{eq:susceptibilities}
\end{eqnarray}
where ${\cal T}$ is the time ordered operator,
$n(\tau)=\sum_{\alpha\sigma} n_{\alpha\sigma}(\tau)$,
$n_{\sigma(\alpha)}(\tau)=\sum_{\alpha(\sigma)}
n_{\alpha\sigma}(\tau)$, and $\tau$ is an imaginary time.

With the knowledge of the single-particle energy spectrum, the optical
conductivity $\sigma_{xx}(\omega)$ can be calculated in the
local approximation \cite{Khurana,Rozenberg,Jarrell}. In terms of the
single-particle spectral weight $A({\bf k},\omega)$,
$\sigma_{xx}(\omega)$ is
\begin{eqnarray}
    \sigma_{xx}(\omega)&=&\frac{e^{2}\pi}{\Omega}\int_{-\infty}^{\infty}
          d\varepsilon \frac{f(\varepsilon)-f(\varepsilon+\omega)}{\omega}
  \nonumber\\
       &&\times\frac{1}{N}\sum_{\bf k \sigma}(
         \frac{\partial\varepsilon_{\bf k}}{\partial k_{x}} )^{2}
          A({\bf k},\varepsilon)A({\bf k},\varepsilon+\omega)
\end{eqnarray}
where $e$ is the electron charge, $\Omega$ is the volume of the
lattice, and $e^{2}\pi/\Omega$ is the unit of the conductivity. The
disappearance of the Drude peak could indicate the occurrence of
MIT, and the optical conductivity peaks provide many information of
the charge excitation between subbands of the systems.

\section{RESULTS AND DISCUSSION}
\label{secmodel}

 As reported in the literature, the Hund's coupling plays an important
role in controlling the Mott transition, and the nature of the Mott
transition in the two-orbital system on the symmetric Bethe lattice has
been controversial \cite{Inaba,Pruschke05}.
In the present study, we investigate the properties of the Mott
transition on the asymmetric triangular lattice so as to resolve
the controversial results.

%%%%%%%%%%%%%%%%%%%%%%%%%%%%%%%%%%%%%%%%%%%%%%%%%%%%%%%%%%%%%%%%%%%%%%%%%%%
\begin{figure}[tp]
\vglue -0.6cm \scalebox{1.150}[1.15]
{\epsfig{figure=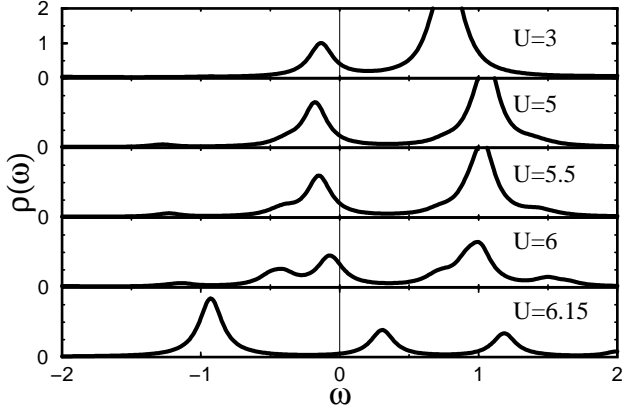,width=5.0cm,angle=270.0}}
\caption{Evolution of density of states (DOS) $\varrho(\omega)$ with
intra-orbital Coulomb
interaction $U$ in two-orbital Hubbard model on triangular lattice.
From top to bottom, U=3, 5, 5.5, 6 and 6.15; $J=U/4$, and $\beta$=16.0}
\label{fig:fig1}
\end{figure}
%%%%%%%%%%%%%%%%%%%%%%%%%%%%%%%%%%%%%%%%%%%%%%%%%%%%%%%%%%%%%%%%%%%%%%%%%%%

Remarkably different from the Bethe lattice, the density of states
(DOS) of quasiparticles on the triangular lattice is asymmetric, as
shown in Fig. 1, where we also present the evolution of the DOS with
the increase of the intra-orbital Coulomb interaction. For $J=U/4$,
it is clearly seen that the Mott transition has already occurred at
$U\backsimeq6.15$. A detailed numerical calculation shows that the
critical value of the MIT is U$_{c}$=6.12.
For other finite $J$, the dependence of DOS on the Coulomb interaction
strength $U$ exhibits similar tendency.
With the increase of the Hund's coupling $J$, the critical points of
the Mott transition occur at 18.2, 16.8, 6.12 and 5.85 for $J$=0.0,
0.01U, $U$/4, and $U$/3, respectively. The tendency of $U_{c}$
substantially decreasing with the increase of the Hund's coupling on
the present 2-dimensional triangular lattice is consistent with the
previous results on the Bethe lattice.
It is very interesting that for various $J$, the critical value,
$U_{C}$, of the MIT on the triangular lattice is about twice larger
than that on the Bethe lattice. This may arise from two facts: one
is from the spin frustration and fluctuation effect on the
triangular lattice; another is from that the orbital fluctuations in
the two-orbital system enhance the metallic character, leading to a
large critical value, U$_{c}$.
%
% Thus it is more favorable of metallic phase in the ground state.
%
%     J  =   0,   0.01U,    U/4,   U/3
%  U_tri = 18.2,  16.8,    6.12,   5.8
%  U_Bet =  9.24,  7.3,    3.5,    --
%
% On the other hand, we do not find the formation of the pseudogap
% state on the triangular lattice, which is different from the cases
% on the square lattice.
%%
% This may be related to the suppression of the magnetic instabilities in
% the frustrated systems, in contrast to the situation in the square
% lattice, where the quasiparticle states are strongly suppressed and a
% pseudogap state forms\cite{Carter}.
%

%%%%%%%%%%%%%%%%%%%%%%%%%%%%%%%%%%%%%%%%%%%%%%%%%%%%%%%%%%%%%%%%%%%%%%%%%%%
\begin{figure}[tp]
\vglue -0.6cm \scalebox{1.150}[1.15]
{\epsfig{figure=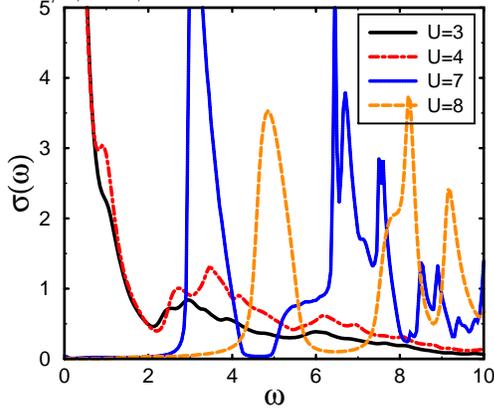,width=5.0cm,angle=270.0}} \caption{(Color
online) Dependence of optical conductivity on frequency in
two-orbital Hubbard model on triangular lattice. The intra-orbital
interactions, $U$=3, 4, 7, and 8; $J=U/4$, the other parameters are
the same as Fig. 1. }
\label{fig:fig2}
\end{figure}
%%%%%%%%%%%%%%%%%%%%%%%%%%%%%%%%%%%%%%%%%%%%%%%%%%%%%%%%%%%%%%%%%%%%%%%%%%%

The optical conductivity also exhibits signatures of the MIT. It is
interesting that how the optical conductivity evolves with the
Coulomb interactions in the two-orbital Hubbard model on the
triangular lattice.
Compared with that of the single-orbital Hubbard model, the optical
conductivity of the two-band Hubbard model is more complicated and
exhibits multi-peak structure, as seen in Fig. 2.
When the Coulomb interaction $U$ is smaller than the critical value
$U_{C}$, the Drude peak and the charge excitation peaks exist at the
same time. The multi-peak charge excitation structure in the present
system significantly differs from the single-peak structure of the
single-orbital Hubbard model \cite{Rozenberg}. The peaks at
$\omega=3.0\thicksim3.5$ and $\omega=6.0\thicksim6.5$ come from the
excitation between different Hubbard subbands below and above the
Fermi surface.
With the increasing of the Coulomb interaction, the intervals of
these Hubbard subbands become larger and larger, and the
charge-excitation peaks move to the high frequency, as seen in Fig.
2.  Since the bandwidths of the two orbitals are identical, no
orbital selective Mott transition is observed.
As $U>$3, we observe a small low-energy mid-peak at $\omega\thicksim 1.0$.
%
% which can be understood from the viewpoint
% of the strong renormalization near the critical point.
%
Such a mid-peak may contribute from the quasiparticle peaks near the
Fermi level, as seen in the DOS near E$_{F}$ in Fig. 1. The
excitation between the renormalized quasiparticle peaks and the
Hubband subbands close to E$_{F}$ gives rise to this small mid-peak.
When the Coulomb interaction $U$ is greater than the critical
interaction $U_{C}$, the Drude peak and the small mid-peak
disappear. Subsequently, the system enters an insulating phase, as
shown in Fig. 2. The insulating gap becomes more and more wide with
the increase of the Coulomb interaction.
Up to date, the optical conductivity experiment on the compounds
with two orbitals and triangular lattice is not available, we
anticipate the corresponding experimental results can be done in
near future.

%%%%%%%%%%%%%%%%%%%%%%%%%%%%%%%%%%%%%%%%%%%%%%%%%%%%%%%%%%%%%%%%%%%%%%%%%%%
\begin{figure}[tp]
\vglue -0.6cm \scalebox{1.150}[1.15]
{\epsfig{figure=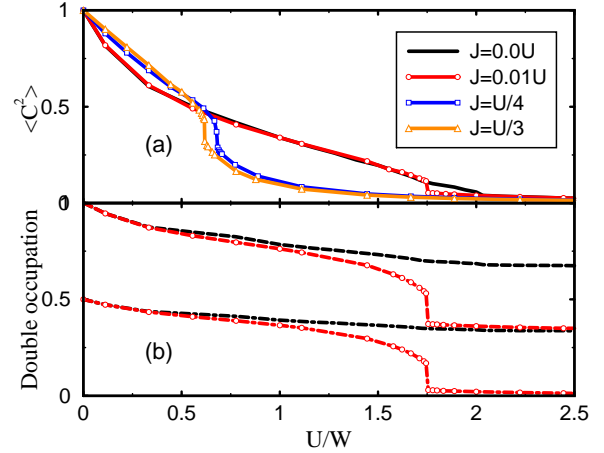,width=5.0cm,angle=270.0}} \caption{(Color
online) (a). Local squared moment of charge as a function of
interaction strength $U/W$ for $J$=0.0, 0.01U, U/4 and U/3. (b).
Dependence of double occupations $<n_{\uparrow}n_{\downarrow}>$
(dashed line) and $<n_{1}n_{2}>$ (dot-dashed line) on U for $J=0.0$
and $J=0.01U$ (circle). }
\label{fig:fig3}
\end{figure}
%%%%%%%%%%%%%%%%%%%%%%%%%%%%%%%%%%%%%%%%%%%%%%%%%%%%%%%%%%%%%%%%%%%%%%%%%%%

As known from the earlier literature, the Hund's coupling J plays a
key role in controlling the nature of the Mott transition in the
two-orbital Hubbard model on the Bethe and the hyper-cubic lattices.
On the present triangular lattice, we investigate the local squared
moment of charge and the double occupations
$<n_{\uparrow}n_{\downarrow}>$ (the dashed lines) and $<n_{1}n_{2}>$
(the dot-dashed lines) on the intra-orbital Coulomb interaction at
different Hund's coupling $J$, as shown in Fig. 3.
At $J/U=0$, as seen in Fig. 3a and Fig. 3b, there is no singular
jump in $<C^{2}>$, $<n_{\uparrow}n_{\downarrow}>$ and $<n_{1}n_{2}>$
in this triangular frustrated system, which implies that the Mott
transition in the two-orbital system with equal bandwidths is the
second order.
On the other hand, when the Hund's coupling J is introduced, there
are discontinuous jumps in $<C^{2}>$ and the double occupancies, as
shown in Fig. 3a and Fig. 3b, showing that the Mott transition in
the two-orbital Hubbard model is first-order. This result is
consistent with what obtained via the Gutzwiller method
\cite{Bunemann:Gutzwiller} on the infinite-dimensional hyper-cubic
lattice.
On the Bethe lattice, the jump of $<S_{z}^{2}>$ is obscure and hard
to distinguish whether the MIT is first- or second-order
\cite{Pruschke05,Inaba}.
While on the present triangular lattice, the jumps of the local
squared moments of charge $<C^{2}>$ is very obvious for all $J\ne0$.
Therefore the MIT in the half-filled two-orbital Hubbard model on
triangular lattice is first-order for all finite J situations.
%
% , even in the physical limit $J=U/3$ or $J=U\prime$.
%
These results also agree with those of Inaba $et$ $al.$ by the DMFT
with self-energy functional approach \cite{Inaba}. However, at
$J=U/4$, our result is contrast to that of Pruschke $et$ $al.$ which
is obtained by the DMFT with the numerical renormalization group
\cite{Pruschke05}.
%
% As known in Ref.[15] and [28], for large J, it is hard to judge the
% order of the MIT from the discontinuity of the local square moment
% of spin, $<S_{z}^{2}>$. Rather, one could easily judge the order of
% the MIT in terms of $<C^{2}>$, as we have seen in Fig.3, implying
%
This shows that $<C^{2}>$ is a proper measure to the MIT in the
large J situation.

To understand the nature of the Mott transition more clearly, we
also calculate the local orbital and spin squared moments of
$<T_{z}^{2}>$ and $<S_{z}^{2}>$, as shown in Fig. 4.
%
%%%%%%%%%%%%%%%%%%%%%%%%%%%%%%%%%%%%%%%%%%%%%%%%%%%%%%%%%%%%%%%%%%%%%%%%%%%
\begin{figure}[tp]
\vglue -0.6cm \scalebox{1.150}[1.15]
{\epsfig{figure=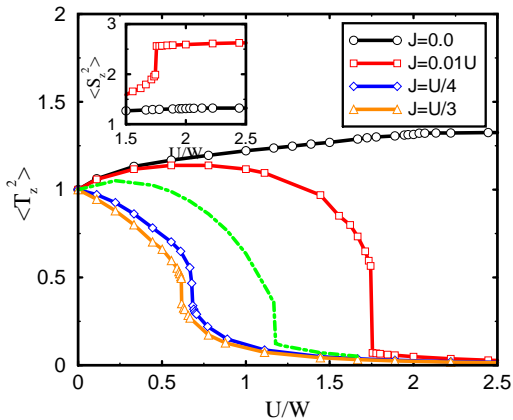,width=5.0cm,angle=270.0}}
\caption{(Color online) Local orbital squared moment vs ratio of intra-orbital
   Coulomb interaction $U$ over bandwidth $W$ for different $J$.
   The green dot-dashed line is for the $SU(2)$$\otimes$$SU(2)$
   system with J$^{\prime}$=0, J=0.1 and U=U$^{\prime}$+J.
   The local spin squared moment vs $U/W$ is shown for $J=0.0$ and
   $J=0.01U$ in the inset.}
\label{fig:fig4}
\end{figure}
%%%%%%%%%%%%%%%%%%%%%%%%%%%%%%%%%%%%%%%%%%%%%%%%%%%%%%%%%%%%%%%%%%%%%%%%%%%
%
In the metallic limit of U=0, the local squared moments
$<T_{z}^{2}>=<S_{z}^{2}>=1$; and in the insulating and strongly
correlated regime, $<T_{z}^{2}>=<S_{z}^{2}>=4/3$ for $J=0$, and
$<T_{z}^{2}>$=0 and $<S_{z}^{2}>=8/3$ for all finite $J$, which are
in agreement with the linearized DMFT results \cite{Ono03}.
%
% The little difference of both is attributed to finite temperature in
% this calculation and local fluctuations are neglected in the L-DMFT for
% the insulating phase.
%
As seen in Fig. 4, the local squared moments of spin and orbital are
continuous at J=0, showing that the MIT is the second order, in
agreement with the result from the local square moment of charge.
Further, as seen in Fig. 4, for various finite J with J=0.01U,
U$/$4, and U$/$3, the discontinuous jumps of the local squared
moments of orbital and spin also demonstrate that the MIT is
first-order, consistent with the preceding results.
Therefore, combining the local squared moment of charge, $<C^{2}>$,
and those of spin and orbital, $<S_{z}^{2}>$ and $<T_{z}^{2}>$, one
can measure the order of the MIT over all of the Hund's coupling J.

%%%%%%%%%%%%%%%%%%%%%%%%%%%%%%%%%%%%%%%%%%%%%%%%%%%%%%%%%%%%%%%%%%%%%%%%%%%
\begin{figure}[tp]
\vglue -0.6cm \scalebox{1.150}[1.15]
{\epsfig{figure=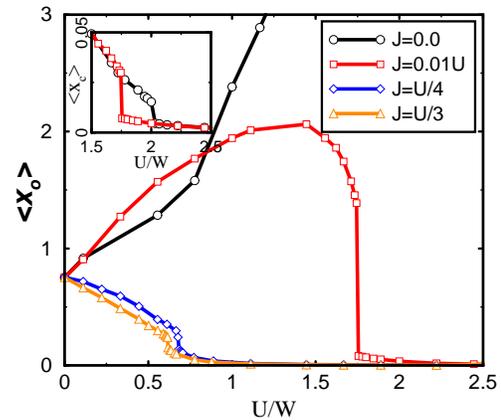,width=5.0cm,angle=270.0}}
\caption{(Color online) Local orbital susceptibility vs ratio of Coulomb interaction
$U$ over bandwidth $W$ for $J=0.0$, $0.01U$, $U/4$ and $U/3$. The inset
shows local charge susceptibility in the case of $J=0.0$ and $J=0.01U$,
respectively.
}
\label{fig:fig5}
\end{figure}
%%%%%%%%%%%%%%%%%%%%%%%%%%%%%%%%%%%%%%%%%%%%%%%%%%%%%%%%%%%%%%%%%%%%%%%%%%%

Consistent with the behaviors of the local squared moments, the
divergence of the local orbital susceptibility near $U_{c}$ in Fig.
5, together with the discontinuous jump of the local charge
susceptibility in the inset in Fig. 5, clearly shows that the MIT on
the triangular lattice is the second order at $J=0$.
Since the MIT in the present system with finite J is the first
order, the local orbital and charge susceptibilities also exhibit
discontinuities.
%
% the local orbital (charge) susceptibility in Fig. 5
% does not reach zero in the insulator phase due to the finite
% temperature effect,
%
It is worthy of noticing that due to the frustration effect on the
present triangular lattice, the local orbital susceptibility in the
system with $J=0.01U$ is suppressed near the MIT critical point;
meanwhile, such suppression is observed only for $J=0.03U$ on the
Bethe lattice \cite{Inaba}. Similar behavior is also observed in the
local squared moment of orbital.

Compared with the single-orbital Hubbard model on the triangular
lattice, the critical value $U_{c}$ of the two orbital model much
larger than that in the single-orbital model\cite{Aryanpour} is
mainly due to the orbital fluctuations.
The physical origin of the different order in the MIT systems with
finite J and J=0 is still a puzzle. B{\"{u}}nemann {et al.}
attributed it to the presence of multiple atomic energy scales in
the two-orbital Hubbard model. This argument may be not true since
there does exist more than one atomic energy scale except U in the
single-orbital Hubbard model. To resolve this puzzle, we suggest
that the order of MIT in the strongly correlated Hubbard model may
depend on the symmetry of the systems. At J=0, the spin-orbital
coupling system is SU(4) symmetric; on the other hand, the
rotational symmetry of the orbital space is usually broken for
finite J. Even if the rotational symmetry of the orbitals exists,
i.e. $U = U^{\prime} + J$ and the inter-orbital Hund's coupling
$J^{\prime} = 0$, the symmetry of the system is SU(2)$\otimes$SU(2).
However, we find that the phase transition in such a system is still
the first order, as seen in the green curve (dot-dashed line) in
Fig. 4.
%
% The low symmetry in the presence of finite J leads to the existence
% of multiple degenerate points in the free energy, hence to the first-order
% phase transitions.
%
% see the red curve in Fig. 4.
%
We also notice that in the two-orbital Hubbard model with the same
bandwidths, the OSMT is excluded. It is interesting to ask what the
order of the OSMT is in the two-orbital triangular Hubbard models
with different bandwidths, which deserves further study.

One notes that in the two-dimensional triangular spin systems, the
geometric frustration is considerable in the strong correlation
regime, so the spatial correlations and fluctuations of spins may be
important. In this situation, the approximation and precision of the
present single-site DMFT approach should be carefully justified.
Fortunately, when we constrain the discussion in the paramagnetic
phases, the precision of such an approximation is well controlled.
This has been demonstrated for the single-orbital Hubbard model in
the triangular lattice by several authors \cite{Aryanpour,Merino}.
Aryanpour {\it et al.}\cite{Aryanpour} and Merino {\it et
al.}\cite{Merino} have shown that the results of the single-orbital
Hubbard model obtained by the single-site DMFT approach are
consistent with those by other methods. And the transport properties
of the 2-dimensional triangular Hubbard model within the single-site
DMFT agrees with the experimental results of the organic
compound\cite{Limelette}.
On the contrary, such a method is failed when it is applied for the
two-dimensional square lattice. This arises from the fact that in
the 2-dimensional triangular lattice, the spatial antiferromagnetic
correlation is greatly suppressed by the geometric frustration, as
pointed out by Aryanpour {\it et al.}\cite{Aryanpour} and Merino
{\it et al.}\cite{Merino}. Another reason is that the coordinate
number of the triangular lattice is considerably larger than that of
the square lattice.

On the other hand, it is highly desirable to extend the present
single-site DMFT approach to the cluster or cellular DMFT approach
so as to well incorporate the spatial fluctuation and the intersite
correlation, as developed by many authors for the single-orbital
models in recent years\cite{Haule,Carter,Laad,Kyung,Zhang}. However,
such an extension to the multi-orbital model meets difficulty since
it goes beyond the ability of the high-performance computing
resources available. And we anticipate that the cluster extension
will not qualitatively alternate our conclusions.

\section{Conclusions}
\label{seccon}

By using the exact-diagonalization DMFT approach, we have
demonstrated that the Hund's coupling J leads to a first-order
metal-insulator transition in the two-orbital Hubbard model with the
degenerate bandwidths in the triangular lattice. The discontinuities
of the local squared moments of the charge, spin and orbital show
that the first-order metal-insulator transition occurs not only in
the small J region, but also in the large J region. Such distinct
behaviors of the systems with finite J and J=0 are attributed to the
lowering of the symmetry of the systems. The multi-peak structure in
the optical conductivity of the two-orbital Hubbard model arises
from the charge excitation among more than two Hubbard subbands.

\acknowledgments

This work was supported by the NSFC of China no.90303013, the BaiRen
Project and the Knowledge Innovation Program of Chinese Academy of
Sciences. Part of the calculations were performed in Center for
Computational Science of CASHIPS and the Shanghai Supercomputer
Center.


\begin{thebibliography}{}

\bibitem{MIT}
M. Imada, A. Fujimori and Y. Tokura, Rev. Mod. Phys. {\bf 70},
1039 (1998).

\bibitem{Georges}
For a review see, A. Georges, G. Kotliar, W. Krauth and M. J.
Rozenberg, Rev. Mod. Phys. {\bf 68}, 13 (1996), and some references
thereafter.
%
%\bibitem{MerinoMcKenzie}
%J. Merino and R. H. McKenzie, Phys. Rev. B {\bf 61}, 7996 (2000).

\bibitem{Kotliar}
G. Kotliar and D. Volhardt, Physics. Today. {\bf 57}, 53 (2004).

\bibitem{Bulla}
R. Bulla, T. A. Costi and D. Vollhardt, Phys. Rev. B {\bf 64},
045103 (2001).

\bibitem{Bulla99}
R. Bulla, Phys. Rev. Lett. {\bf 83}, 136 (1999).

\bibitem{Onoda}
S. Onoda and M. Imada, J. Mag. Mag. Mat. 272-276 Suppl. 1, E275 (2004).

\bibitem{Kugel}
 K. I. Kugel and D. I. Khomskii, {Sov. Phys.} JETP, {\bf 37}, 725 (1973);
 Y. Tokura and N. Nagaosa, {Science} {\bf 288}, 462 (2000).

\bibitem{Anisimov}
V. I. Anisimov, I. A. Nekrasov, D. E. Kondakov, T. M. Rice and
M.Sigrist, Eur. Phys. J. B {\bf 25}, 191 (2002).

\bibitem{Koga04}
A. Koga, N. Kawakami, T. M. Rice and M. Sigrist, Phys. Rev. Lett.
{\bf 92}, 216402 (2004).

\bibitem{Liebsch}
A. Liebsch, Europhys. Lett. {\bf 63}, 97 (2003).

\bibitem{Liebsch03}
A. Liebsch, Phys. Rev. Lett. {\bf 91}, 226401 (2003).

\bibitem{Liebsch04}
A. Liebsch, Phys. Rev. B. {\bf 70}, 165103 (2004).

%CCCCCCCCCCCCCCCCCCCCCCCCCCCCCCCCCCCCCCCCCCCCCCCCCCCCCCCCCCCCCCCCCCCCC

\bibitem{Inaba}
K. Inaba and A. Koga, Phys. Rev. B {\bf 73}, 155106 (2006);
K. Inaba and A. Koga, J. Phys. Soc. Jpn. {\bf 76}, 094712 (2007).

\bibitem{Inaba05}
K. Inaba, A. Koga, S.-I Suga and N. Kawakami, Phys. Rev. B {\bf 72},
085112 (2005).

\bibitem{Pruschke05}
Th. Pruschke and R. Bulla, Eur. Phys. J. B {\bf 44}, 217 (2005).

\bibitem{Bunemann:Gutzwiller}
J. B{\"{u}}nemann and W. Weber, Phys. Rev. B {\bf 55}, 4011 (1997)
 ; J. B{\"{u}}nemann, W. Weber and F. Gebhard, {\it ibid} {\bf 57},
  6896 (1998).

\bibitem{kanoda} K. Kanoda, Physica C {\bf 282-287}, 299 (1997);
K. Kanoda, Hyperfine Interact. {\bf 104}, 235 (1997).

\bibitem{mckenzie} R. H. McKenzie, Science, {\bf 278}, 820 (1997).

\bibitem{Petit}
L. Petit, G. M. Stocks, T. Egami, Z. Szotek and W. M. Temmerman,
Phys. Rev. Lett. {\bf97}, 146405 (2006).

\bibitem{Coldea}
E. Wawrzynska, R. Coldea, E. M. Wheeler, I. I. Mazin, M. D.
Johannes, T. S{\"{o}}rgel, M. Jansen, R. M. Ibberson and P. G.
Radaelli, Phys. Rev. Lett. {\bf99}, 157204 (2007).

\bibitem{Aryanpour}
K. Aryanpour, W. E. Pickett and R. T. Scalettar, Phys. Rev. B {\bf
74}, 085117 (2006).

\bibitem{Song05}
Y. Song and L.-J. Zou, Phys. Rev. B {\bf 72}, 085114 (2005).

\bibitem{Castellani}
C. Castellani, C. R. Natoli and J. Ranninger, Phys. Rev. B. {\bf18},
4945 (1978).

\bibitem{Liebsch05}
A. Liebsch, Phys. Rev. Lett. {\bf 95}, 116402 (2005).

\bibitem{Demchenko}
D. O. Demchenko, A. V. Joura and J. K. Freericks, Phys. Rev. Lett.
{\bf92}, 216401 (2004).

\bibitem{Rozenberg}
M. J. Rozenberg, G. Kotliar, H. Kajueter, G. A. Thomas, D. H.
Rapkine, J. M. Honig and P. Metcalf , Phys. Rev. Lett {\bf 75}, 105
(1995).

\bibitem{Khurana}
A. Khurana, Phys. Rev. lett {\bf 64}, 1990 (1990).

\bibitem{Jarrell}
M. Jarrell, J. K. Freericks and Th. Pruschke, Phys. Rev. B {\bf 51},
11704 (1995).

\bibitem{Ono03}
Y. Ono, M. Potthoff and R. Bulla, Phys. Rev. B {\bf 67}, 035119
(2003).

\bibitem{Merino} J. Merino, B. J. Powell and R. H. McKenzie, Phys. Rev. B.
{\bf73}, 235107 (2006).

\bibitem{Limelette} P. Limelette, P. Wzietek, S. Florens, A. Georges, T. A. Costi,
 C. Pasquier, D. Jerome, C. Meziere and P. Batail, Phys. Rev. Lett.
{\bf91}, 016401 (2003).

\bibitem{Carter} E. C. Carter and A. J. Schofield, Phys. Rev. B {\bf70},
045107 (2004).

\bibitem{Haule} K. Haule and G. Kotliar, Phys. Rev. B {\bf76},
104509 (2007).

\bibitem{Laad} M. S. Laad and L. Craco, cond-mat/0701585.


\bibitem{Kyung} B. Kyung, G. Kotliar and A.-M. S. Tremblay, Phys. Rev. B {\bf73},
205106 (2006).

\bibitem{Zhang} Y. Z. Zhang and Masatoshi Imada, Phys. Rev. B {\bf76},
045108 (2007).


%CCCCCCCCCCCCCCCCCCCCCCCCCCCCCCCCCCCCCCCCCCCCCCCCCCCCCCCCCCCCCCCCCCCCCCCCCCCCC

%\bibitem{Parcollet}
%O. Parcollet and G. Biroli, Phys. Rev. Lett. {\bf 92}, 226402
%(2004).

%\bibitem{Ohashi}
%T. Ohashi, N. Kawakami and H. Tsunetsugu, Phys. Rev. Lett. {\bf 97},
%066401 (2006).


%\bibitem{Pietig}
% R. Pietig,R, Bulla, S. Blawid, Phys. Rev. lett {\bf 20}, 4046 (1999).

%\bibitem{Takimoto}
%T. Takimoto, Phys. Rev. B {\bf 62}, R14641 (2000).

%\bibitem{takada} K. Takada, H. Sakurai, E. Takayama-Muromachi, F. Izumi,
%R. A. Dilanian and T. Sasaki, Nature (London) {\bf 422}, 53 (2003).

%\bibitem{okamoto}
% Y. Okamoto, M. Nohara, H. Aruga-Katori and H. Takagi, cond-mat/0705282.

%\bibitem{tennant}
%cond-mat/0507040

%\bibitem{Nozieres}
%See, for example, P. Nozieres, {\it Theory of interacting Fermi
%systems}, Addison-Wesley, (1997).

\end{thebibliography}
\end{document}